\documentclass[aps,prl,twocolumn,showpacs,preprintnumbers,floatfix,reprint]{revtex4}
\usepackage{amsfonts}
\usepackage{amsmath}
\usepackage{amssymb}
\usepackage{graphicx}
\usepackage{dcolumn}
\usepackage{color}

\begin{document}

\title{Exact multi-soliton solutions in the four dimensional Skyrme model}
\author{Fabrizio Canfora$^{1,2}$, Francisco Correa$^{1}$, Jorge Zanelli$^{1,2}$\\
$^{1}$\textit{Centro de Estudios Cient\'{\i}ficos (CECS), Arturo Prat 514, Valdivia, Chile.}\\$^{2}$\textit{Universidad Andres Bello, Av. Rep\'ublica 440, Santiago, Chile.}\\{\small canfora@cecs.cl, correa@cecs.cl, z@cecs.cl}}

\pacs{
12.39.Dc, 11.27.+d, 05.45.Yv, 11.10.Lm }

\begin{abstract}
Exact analytic solutions of the Skyrme model defined on a spherically symmetric $R^{(1,1)} \times S^2$ geometry, chosen to mimic finite volume effects, are presented. The static and spherically symmetric configurations have non-trivial winding number and finite soliton mass. These configurations possess an extra topological charge, allowing for a novel BPS bound which can be saturated, unlike what happens in the standard case. Such solutions include exact multi-Skyrmions of arbitrary winding number, composed by interacting elementary Skyrmions. The values of the coupling constants and the compression modulus are found in good agreement with experiments by fitting the masses $M_N$ and $M_\Delta$.
\end{abstract}

\maketitle

%\address{$^{1}$\textit{Centro de Estudios Cient\'{\i}ficos (CECS), Arturo Prat 514, Valdivia, Chile.}\\
%$^{2}$\textit{Universidad Andres Bello, Av. Rep\'ublica 440, Santiago, Chile.}\\
%{\small canfora@cecs.cl, correa@cecs.cl, z@cecs.cl}}

Skyrme's theory \cite{skyrme} is an important model in physics due to its wide range of applications. A remarkable property of the Skyrme term is that is supports static finite energy soliton solutions-\textit{Skyrmions} --representing Fermionic degrees of freedom describing nucleons \cite{multis2,manton,witten,footnote1,ANW,guada}. One of the most compelling results in this area was the identification of the winding number of the Skyrmion with the baryon number in particle physics. The relevance of Skyrme theory in other areas such as astrophysics \cite{astroSkyrme}, Bose-Einstein condensates  \cite{useful1}, nematic liquids \cite{useful2}, magnetic structures \cite{useful4} and condensed matter physics in general \cite{useful7}, is well recognized by now. Furthermore, in the context of AdS/CFT it has been shown that the Skyrme model appears in a very natural way \cite{sakaisugimoto}.

In spite of all this, until very recently no exact analytic solutions of the Skyrme model with non-trivial topological charges were known. One of the reasons is that the Skyrme-BPS bound on the energy cannot be saturated for non-trivial spherically symmetric configurations \cite{footnote2}. Nevertheless, many rigorous results about Skyrmions dynamics have been derived, see for instance \cite{moduli, skyrmonopole, rational}.

The action of the $SU(2)$ Skyrme system in four dimensional spacetime is 
\begin{align}
\!\!\!\!\! S_{\mathrm{Skyrme}} & =\frac{K}{2}\int d^4 x \sqrt{-g}\,\mathrm{Tr}\left(\frac{1}{2}R^\mu R_\mu +\frac{\lambda}{16} F_{\mu\nu}F^{\mu\nu}\right), \!\!  \label{skyrmaction}
\end{align}
where $R_\mu =U^{-1}\nabla_\mu U=R_\mu ^i t_i$ and $F_{\mu\nu}:=\left[R_\mu,R_\nu \right]$. Here $t_i$ are the $SU(2)$ generators and we set the units $\hbar=c=1$. The coupling constants $K>0$ and $\lambda>0$ are fixed by comparison with experimental data \cite{ANW}. The presence of the first term of the Skyrme action (\ref{skyrmaction}), is mandatory to describe pions while the second is the only covariant term leading to second order field equations in time which supports the existence of Skyrmions in four dimensions.

In the present paper, exact spherically symmetric solutions of the Skyrme model with both a non-trivial winding number and a finite soliton mass (topological charge) are presented. Using the formalism introduced in \cite{canfora1,canfora2,canfora3}, it is shown that although the BPS bound in terms of the winding cannot be saturated, a new topological charge exists that can be saturated corresponding to a different BPS bound. The baryon number is the homotopy of the space into the group. The simplest choice would be to consider the curved background $S^3$ as physical space, as already considered in the pioneering papers \cite{curved1, curved2}. The second natural choice of special sections with integer homotopy into $SU(2)$ is $S^1 \times S^2$ (or $\mathbb{R} \times S^2$). This can be represented by a metric of
the form 
\begin{equation}
ds^2 =-dt^2 +dx^2 + R_0^2 (d\theta^2 + \sin^2\theta\, d\phi^2). \label{metric}
\end{equation}
In simple words, this geometry describes tridimensional cylinders whose sections are $S^{2}$ spheres of area $4\pi R_{0}^{2}$. The physical meaning of $R_{0}$ is that it takes into account finite volume effects. One could put the Skyrme action into, say, a cube. However, this way of proceeding often breaks symmetries. On the other hand, a spherical box of finite radius would lead to difficulties in requiring the Skyrmions approach the identity at the boundary. Therefore, it is much more convenient to choose a metric which at the same time takes into account finite volume effects and keeps the spherical symmetry. We are able construct exact Skyrmions in a finite volume but, instead of putting by hand a cut-off on the coordinates, we leave this task to the geometry. Besides, this geometry is such that the group of the isometries of (\ref{metric}) contains $SO(3)$ as a subgroup and so it includes the spherical symmetry of the Skyrmion in flat space. This fact allows examining how far is the BPS bound from being saturated and to construct an energy bound which can in fact be saturated. This could be of interest both in high energy and solid state physics whose features, after the papers \cite{curved1, curved2}, have not been thoroughly investigated from the analytical viewpoint.

In order to construct the exact solution of the Skyrme model, the following standard parametrization of the $\mathfrak{su}(2)$-valued scalar $U(x^{\mu })$ will be adopted, 
\begin{equation}
U^{\pm 1}(x^{\mu })=Y^{0}(x^{\mu })\mathbf{1}\pm Y^{i}(x^{\mu })t_{i}\ ,\ \
\left( Y^{0}\right) ^{2}+Y^{i}Y_{i}=1\ .  \label{standard1}
\end{equation}
Thus, the hedgehog ansatz describing a spherically symmetric Skyrmion in this background \cite{canfora2} can be written in terms of the unit vectors 
\begin{equation}
Y^{0}=\cos \alpha \ ,\ \ Y^{i}=\widehat{n}^{i}\sin \alpha \ ,\ \ \ \alpha =\alpha (x)\ ,  \label{hedge1}
\end{equation}
and $\widehat{n}^{1}=\sin \theta \cos \phi $, $\widehat{n}^{2}=\sin \theta \sin \phi $ and $\widehat{n}^{3}=\cos \theta $. As it happens for the usual spherically symmetric Skyrmion in flat space, the energy-momentum tensor in the present case also corresponds to a spherically symmetric distribution, in spite of $Y^{0}$'s and $Y^{i}$'s explicit angular dependence. The Skyrme field equations of the action (\ref{skyrmaction}) are 
\begin{equation}
\nabla ^{\mu }R_{\mu }+\frac{\lambda }{4}\nabla ^{\mu }[R^{\nu },F_{\mu \nu}]=0.  \label{nonlinearsigma1}
\end{equation}
Remarkably, for the ansatz (\ref{hedge1}), the Skyrme field equations (\ref{nonlinearsigma1}), reduce to one ordinary differential equation for the Skyrmion profile $\alpha $: 
\begin{equation}
\left( \!1\!+\!\frac{2\lambda }{R_{0}^{2}}\sin ^{2}\alpha \!\right) \alpha^{\prime \prime }-\frac{\sin (2\alpha )}{R_{0}^{2}}\left( \!1\!-\!\lambda \left[ \alpha ^{\prime 2}\!-\!\frac{\sin ^{2}\alpha }{R_{0}^{2}}\right] \!\right) =0\ ,  \label{hedge3}
\end{equation}
where $\alpha ^{\prime }=d\alpha /dx$. This equation reduces to the sine-Gordon form for $\lambda=0$, and also when a dimensional reduction in flat 1+1 dimensional spacetime as noted by  Perring and Skyrme \cite{skyrmeperring}. However, it can also be exactly integrated for any $\lambda$ as
\begin{equation}
(\alpha^{\prime})^2=[I +2G(\alpha )][F(\alpha )]^{-1}\ ,
\label{hedge6.1}
\end{equation}
where $I$ is an integration constant and we have defined 
\begin{align}
F(\alpha )& =1+\frac{2\lambda }{R_{0}^{2}}\sin ^{2}\alpha ,  \notag \\
G(\alpha )& =\frac{\sin ^{2}\alpha }{R_{0}^{2}}\left( 1+\frac{\lambda }{2R_{0}^{2}}\sin ^{2}\alpha \right) .
\end{align}
In this way, Equation (\ref{hedge3}) has been reduced to a first integral (quadrature). The expression (\ref{hedge6.1}) is also appropriate to discuss the boundary conditions and, in particular, the possibility of having non-trivial winding number (see below) \cite{footnote3}.

The energy density is 
\begin{eqnarray}
T_{00} &=&\frac{K}{2}\left[ F(\alpha )(\alpha^{\prime 2}+2G(\alpha ))\right] \qquad \qquad \,  \label{hedge4} \\
&=&\frac{KF}{2}\left[\alpha^{\prime }\mp \left(\frac{2G}{F}\right)^{1/2} \right]^2 \pm \sqrt{2}K\left[ FG\right] ^{1/2}\alpha' \,,  \notag
\end{eqnarray}
from which an inequality for the total energy is found, 
\begin{align}
E_{tot}& =4\pi R_{0}^{2}\int T_{00}dx\geq \left\vert Q\right\vert \ , \label{bpsb1} \\
Q& =\sqrt{32}\pi R_{0}^{2}K\int \left( \left[ F(\alpha )G(\alpha )\right]^{1/2}\frac{d\alpha }{dx}\right) dx\,.  \label{bpsb2}
\end{align}
Clearly, $Q$ is a boundary term and is therefore invariant under continuous deformations of the fields in the bulk, as suits a topological invariant. This inequality is saturated if $\alpha $ satisfies the first-order differential equation 
\begin{equation}
\alpha ' =\pm \sqrt{2G(\alpha )/F(\alpha )}\, , \label{bpsb3}
\end{equation}
which coincides with (\ref{hedge6.1}) for $I=0$.

The winding number for the generalized hedgehog ansatz in (\ref{hedge1}) is 
\begin{align}
W&=-\frac{1}{24\pi^{2}}\int\epsilon^{ijk}Tr (U^{-1}\partial_{i}U)(U^{-1}\partial_{j}U) (U^{-1}\partial_{k}U)  \notag \\
   % &= \frac{2}{\pi}\int(\alpha^{\prime 2 }\sin^2 \alpha) dx 
    &= \frac{2}{\pi}\int_{\alpha(x_1)}^{\alpha(x_2)}\sin^2 \alpha \, d\alpha ,  \label{winding}
\end{align}
where $(x_1,x_2)$ are the limits in the spatial direction, that can be taken as $(-L/2,L/2)$ or $(-\infty,\infty)$. The winding number takes integer values $n$ for boundary conditions 
\begin{equation}
\alpha(x_2) - \alpha(x_1) = n\pi\, ,\ \ n \in \mathbb{Z}\, .
\label{windinteger}
\end{equation}
These boundary conditions are unique in that they ensure $U(x_2)=(-1)^n U(x_1)$, which correspond to bosonic and fermionic states for even and odd $n $, respectively. Moreover, it can be seen that the topological charge $Q$ is bounded from below, $|Q|>|W|$, as it should be \cite{manton}.

Smooth solutions exist in a finite range $(-L/2,L/2)$, for any $n$ satisfying the above boundary conditions. Such solutions depend on the value of $I$, as can be seen by integrating (\ref{hedge6.1}),
\begin{equation}
\frac{L}{n}=\pm \int_{0}^{\pi }\sqrt{\frac{F(\alpha )}{I +2G(\alpha )}}d\alpha   \label{ln}
\end{equation}
When the bound is saturated ($I=0$), the $x$ coordinate must range from $-\infty $ to $\infty$ and the winding number is $n=\pm 1$, so that the ground state is Fermionic. The profile of $\alpha$ and the corresponding energy density $T_{00}$, are depicted in Fig. 1. The proof that elementary Skyrmions are Fermions is topological in nature \cite{witten}. The same argument applies in the present case, where the Fermionic nature elementary Skyrmions is also supported by the repulsive interactions appearing in the multi-Skyrmion configurations (see below).
\begin{figure}[h!]
\centering
\includegraphics[scale=.65]{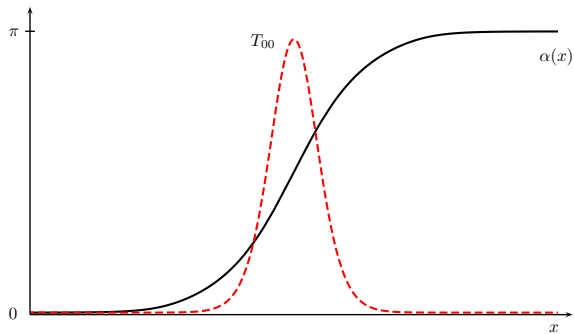}\label{fig1}
\caption{Profile of $\protect\alpha$ (solid line) and $T_{00}$ (dashed line) for the ground state $I=0$. The energy density is rescaled.}
\end{figure}

Eq. (\ref{ln}) can be seen as giving $I=I(n;L)$. It can be easily seen that for large $n$, $I$ grows as $n^2/ L^2$. Eq. (\ref{hedge6.1}) admits multi-soliton solutions which represent Skyrmions with winding number $n$ as shown in Fig. 2.
\begin{figure}[h]
\label{figure3} \centering
\includegraphics[scale=.65]{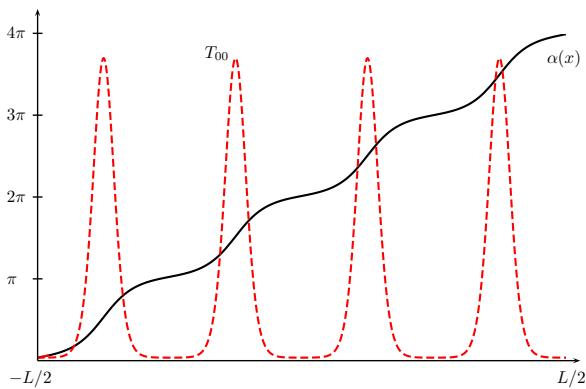}
\caption{Profile of $\protect\alpha $ for $I>0$ with winding number $n=4$. The kink becomes sharper for smaller $R_{0}$.}
\end{figure}
The multi-Skyrmion solutions do not saturate the bound and are not simple linear superpositions of single Skyrmions. They include strong repulsive interactions among the elementary Skyrmions. In particular, in the limit of large $n$ and fixed $L$, the energy of the multi-Skyrmions grows as $n^{2}$, as shown in Fig. 3 (A similar behavior has been found in a modified version of the Skyrme model \cite{AW}). 
\begin{figure}[h]
\centering
\includegraphics[scale=.7]{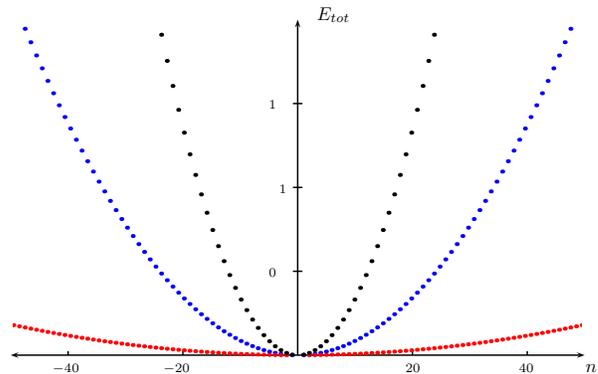}
\caption{$E_{tot}$ (\protect\ref{bpsb1}) as a function of the winding number $n$, for $R_{0}=4$ (upper curve), $R_{0}=2$ and $R_{0}=0.5$ (lower curve) for $L=10$ and $\protect\lambda =0.1$. The plot grows roughly as $n^{2}$. }
\end{figure}
These multi-Skyrmion form regular patterns along $x$, as in Fig. 2,  and for periodic $x$, they look like necklaces of elementary Skyrmions with parallel isospin vector. The tendency of Skyrmions to form ordered arrangements has been investigated numerically in \cite{sut1}. Up until now, analytical studies had only been done for configurations with trivial winding number \cite{canfora3,chen}). 

In closing, the following comments are in order: \newline
\textbf{1.} Within the family of solutions for (\ref{hedge6.1}), the case $I=3/4\lambda$ is exceptional. For this value the profile obeys 
\begin{equation*}
(\alpha^{\prime})^2 =\frac{\frac{3}{4\lambda }+\frac{2}{R_0^2}\sin^2 \alpha +\frac{\lambda }{R_0^4}\sin ^{4}\alpha }{\left(
1+\frac{2\lambda }{R_{0}^{2}}\sin ^{2}\alpha \right)}=\frac{3}{4\lambda}+\frac{1}{2 R_{0}^{2}} \sin ^{2}\alpha \,,
\end{equation*}
which reduces to the sine-Gordon equation after differentiation, $\alpha ^{\prime \prime}-(2R_{0})^{-2}\sin 2\alpha =0$. Note that this reduction is valid for any value of $\lambda$ and it differs from the case with $\lambda=0$, which modifies the action (\ref{skyrmaction}). Thus, the solution can be written as 
\begin{equation*}
\alpha (x)=\mathrm{am}\left( {\textstyle\frac{\sqrt{3}}{2\sqrt{\lambda }}}\,x,i\sqrt{{\textstyle\frac{2\lambda }{3R_{0}^{2}}}}\right) 
\end{equation*}
where $\mathrm{am}(u,k)$ is the Jacobi amplitude function, related to the Jacobi elliptic functions $\mathrm{sn}\,(x,k)$ and $\mathrm{cn}\,(x,k)$ by $\mathrm{sn}\,(x,k)=\sin \mathrm{am}(x,k)$, $\mathrm{cn}(x,m)=\cos \mathrm{am}(x,k)$, where $k$ is the modulus parameter defined in the range $0<k<1$ \cite{elliptic}. From (\ref{ln}), the parameters of the solution are found to be related as $L=\frac{2\sqrt{\lambda}}{\sqrt{3}}\mathbb{F}\left( n\pi ,i\sqrt{{\textstyle\frac{2\lambda}{3R_{0}^{2}}}}\right) $, where $\mathbb{F}(\phi ,k)$ is the incomplete elliptic integral of the first kind whose inverse is the Jacobi amplitude function. In this case the group element $U$ takes the form  
\begin{equation}
U^{\pm 1}(x^{\mu })\!=\!\mathrm{cn}\!\left( {\textstyle\frac{\sqrt{3}}{2\sqrt{\lambda }}}\,x,i\sqrt{{\textstyle\frac{2\lambda }{3R_{0}^{2}}}}\right) \mathbf{1}\,\pm \,\widehat{n}^{j}\,\mathrm{sn}\!\left( {\textstyle\frac{\sqrt{3}}{2\sqrt{\lambda }}} \,x,i\sqrt{{\textstyle\frac{2\lambda }{3R_{0}^{2}}}}\right) \!t_{j}  \notag  \label{contact}
\end{equation}
Note that as required by (\ref{standard1}), the Jacobi elliptic functions satisfy the relation $\mathrm{cn}^2 (x,k) + \mathrm{sn}^2(x,k)=1$ \cite{elliptic}.\newline
\textbf{2.} It is clear that the same construction can also be carried out if one includes a mass term in the Skyrme action. The only change in eq. (\ref{hedge6.1}), is the replacement of the function $G(\alpha )\rightarrow G(\alpha )-m^{2}\cos \alpha $. \newline
\textbf{3.} These multi-Skyrmions live on a curved background which allows to explore finite-size effects. This is an important technical point in order to compute thermodynamic quantities for which the explicit dependence of the total energy of the multi-Skyrmions of winding number $n$ on the volume $V$ is needed. In the present case, eq. (\ref{hedge6.1}) allows to write down the total energy, changing the integration variable in (\ref{bpsb1}) from the $dx$ to $d\alpha $, 
\begin{equation*}
E_{tot}=2\pi KR_{0}^{2}\left(IL+4n\int_{0}^{\pi}\frac{GF^{1/2}}{\sqrt{I+2G}}\,d\alpha \right) \,,
\end{equation*}
which determines its dependence both on the volume and the winding number.
\newline
\textbf{4.} The present setup can be tested by computing some relevant phenomenological observables. For instance, in the collective coordinate approach it is possible to identify the quantum ground state, following the ideas in ref. \cite{ANW}. The eigenvalues of the corresponding quantum Hamiltonian are 
\begin{equation*}
E_{\ell }=M+\frac{1}{8\Lambda }\ell (\ell +2)\,,
\end{equation*}
where $\ell $ is a positive integer, and 
\begin{equation*}
\textstyle M=2\pi KR_{0}^{2}\int_{-\infty }^{\infty }\left[ \alpha ^{\prime
\,2}\!+\!\frac{2\sin ^{2}\alpha }{R_{0}^{2}}\!+\!\lambda \frac{\sin
^{2}\alpha }{R_{0}^{2}}\left( 2\alpha ^{\prime \,2}\!+\!\frac{\sin
^{2}\alpha }{R_{0}^{2}}\right) \right] dx\,,
\end{equation*}%
\begin{equation*}
\Lambda =\frac{8\pi KR_0^2}{3}\int_{-\infty}^{\infty} \sin^2 \alpha \left[1+\lambda \left(\alpha'^2 +\frac{\sin^2 \alpha}{R_0^2}\right) \right] dx\,.
\end{equation*}
In this way, one can fit the neutron and delta masses ($M_{N}=939$ MeV, $M_{\Delta }=1232$ MeV) by choosing $R_0 =0.65$ fm, which yields $ F_{\pi}=$141 MeV, $e=5.45$, and can be compared with \cite{ANW}. These parameters are related to the constants in the action by $F_{\pi }=2\sqrt{K}$ and $K\lambda e^{2}=1$. 

\textbf{5.} Having fixed $R_{0}$ by fitting $M_{\Delta }$ and $M_{N}$, it is now possible to compute the compression modulus, for which the prediction in the Skyrme model is an outstanding problem \cite{ANSSW}. The compression modulus,
\begin{equation}
\kappa=\frac{9V^{2}}{n}\left. \frac{\partial ^{2}E_{tot}}{\partial V^{2}}\right\vert_{T,n}
 \label{compression}
\end{equation}
can be calculated assuming $L$ to be proportional to $R_{0}$. This implies that the volume is proportional to the Baryon number $n$, which is an experimental fact in nuclear physics. In this way, the volume is found to be $V\approx 1.42 n \ \text{fm}^3$, and the compression modulus for $n=100$ approaches the experimental value $\kappa = 230$ MeV \cite{compression,CCZ2}.

\textbf{6.} These results indicate that the idea of describing finite-volume effects by using a metric with finite spatial volume (instead of putting the theory in a box) opens up a new analytic path to study Skyrmion thermodynamics which deserves further investigation. This method not only allows the explicit construction of multi-Skyrmion configurations, but also gives results in reasonable agreement with experiments for $R_0$ of the order of the proton radius. If more support is gained by this model, a natural question would be why is the nuclear phenomenology so well described by the special geometry (\ref{metric}).

\emph{Acknowledgments}. The authors thank E.~Ay\'on-Beato, J.~Oliva and G.~Tallarita for enlightening discussions. This work was partially funded through Fondecyt grants 1120352, 11121651 and 1140155. F. Correa was partially supported by Conicyt grant 79112034. CECs is funded by the Chilean Government through the Centers of Excellence Base Financing Program of Conicyt.


\begin{thebibliography}{99}
\bibitem{footnote2} The only case in which the bound with the winding number can be saturated corresponds to choosing the spatial section as $S^3$ \cite{curved1, curved2}. Here we analyze the case in which the bound cannot be saturated in terms of the winding, which is phenomenologically reasonable.
\bibitem{footnote3} Time-dependent solutions can be easily obtained by applying Lorentz boosts in the $x-t$ plane to $\alpha(x)$, producing Skyrmions moving with constant velocity.
\end{thebibliography}
\end{document}